\documentclass{article}
\usepackage{hyperref,spconf,amsmath,graphicx,booktabs,pifont,xcolor,colortbl,multirow,url,placeins,stfloats,balance,amssymb}

\definecolor{NCPendingRed}{RGB}{170,0,0}
\definecolor{NCPendingBg}{RGB}{255,242,204}

\definecolor{NCOursBg}{HTML}{DFF0DC}

\title{X2Streaming-ASR: Wait When Uncertain, Emit When Ready \\
  for Streaming ASR}
\name{
Zhiwei Lin, Kaiqi Fu, Rime Wen, Zehan Liu, Shawn Qin, Roy Gan, Hao Wang
}
\address{X Square Robot\\[2pt] 
\small{
        \{linzhiwei,wanghao\}@x2robot.com
    }
}

\begin{document}
\ninept
\maketitle



\begin{abstract}
Streaming automatic speech recognition (ASR) for real-time voice agents and
full-duplex dialogue must provide accurate partial transcripts with low
commit latency.
Existing systems commonly use a fixed chunk size, look-ahead, or target delay,
or encourage emissions near estimated acoustic boundaries.
These approaches do not directly optimize how much additional context to use
at each output position under a single-pass, hard-commit constraint.
We propose X2Streaming-ASR, which decomposes streaming recognition into
when to commit and what to commit.
Its three-stage training procedure first establishes streaming recognition
ability, then warm-starts the commit policy with automatically probed
trajectories, and finally refines the policy using character-level,
segment-assigned group-relative rewards for recognition accuracy and latency.
Across ten Chinese and English test sets, X2Streaming-ASR attains the lowest
mean commit latency relative to forced-aligned endpoints,
\(32\)--\(109\,\mathrm{ms}\) on Chinese characters and
\(12\)--\(85\,\mathrm{ms}\) on English words, while recognition accuracy remains
comparable to existing systems.
Code is available at \url{https://github.com/X-Square-Robot/X2Streaming-ASR}.
\end{abstract}


\section{Introduction}

Streaming speech recognition (ASR) requires the system to output text
in real time during speech input, the foundational capability for
real-time captioning and voice interaction.
The main metrics for evaluating streaming ASR are recognition accuracy
(CER/WER) and latency.
This paper focuses on character-level emission latency relative to
forced alignment: how long it takes from a character's acoustic
boundary to the system outputting it.
For cascaded voice agents, the traditional process waits for offline
transcription before starting downstream services.
Some works \cite{zou2026lts} trigger downstream services on streaming
recognition prefixes, reducing the overall system response time.
However, incorrect recognition will compromise subsequent services,
and recognition that arrives too late will delay service startup.
Furthermore, recent work on turn-taking introduces streaming ASR so
that decisions can use semantic information, not just acoustic
information~\cite{yan2026soulx, fu2026x2}.
Therefore, the latency and accuracy of streaming ASR directly determine
whether downstream services and turn-taking can be started as early
and correctly as possible.

Existing work still struggles to jointly optimize accuracy and emission latency, and falls into three categories.
The first approach uses global configuration for streaming recognition, such as fixed chunk decoding and look-ahead or delay\cite{yao2024zipformer,paraformer,an2022cuside,cuside-t,voxtral,DSM}. 
Chunking specifies the decoding unit, while look-ahead and delay determine the amount of future context that is visible.
Neither of these methods determines how much future information to wait for based on historical context and current acoustic information.
The second approach shifts the emission time during training. 
FastEmit \cite{yu2021fastemit} encourages earlier output, whereas minLT \cite{inaguma2020minimum, shinohara2022minimum, wan2026streaming} pulls emissions toward the alignment boundary. Unlike a global delay, they do not wait longer where upcoming context is needed for disambiguation.
The third\cite{machavcek2023turning,liu2020low, shi2026qwen3, xia2026uni} emits an unstable or partial hypothesis and later revises it.
These designs miss the same fact: the amount of future context needed after acoustics boundary is position-dependent.
A position-agnostic knob cannot implement this position-dependent waiting: setting it later delays every position, while setting it earlier exposes every position to the same risk. Treating “emit at the boundary” as the training objective forbids additional wait on hard characters. 
Revision-based methods reduce latency by correcting earlier output, and are therefore not single-pass hard commit. 
In this paper, we consider the setting where each character is emitted only once, with no second-pass correction.

Streaming ASR makes two decisions: when to commit and what to
commit. Supervised learning already handles the second well. The
hard one is the first, which depends on acoustics and context, and
fitting emit times with a supervised target works poorly. On the data
side, the first time a character can be committed is not always its
acoustic boundary, so aligned $t_{\mathrm{end}}$ is only a proxy for
the emit target, not a unique gold time. Teacher-probed
``earliest correct'' times are also imperfect: they inherit the
teacher's language priors and the probe protocol. On the training
side, the cross-entropy loss only fits these labeled times; latency
and accuracy are not in the objective. Waiting slightly longer to
recognize the character correctly, or emitting correctly earlier than
the label, may help the real decision, but both count as deviations
from this loss and are penalized.

Reinforcement learning closes both gaps. First, it needs no
pre-labeled recognizable points: multiple trajectories are sampled
from the same audio, so the same character is committed at different
times. The consequences of waiting and of committing early appear
directly in the reward. Second, it optimizes the true objective: the
reward combines the error count and latency, so the beneficial
deviations that supervised learning would penalize are rewarded here.
We therefore propose X2Streaming-ASR, trained in three stages so the
model waits only where leftover context is still required. Stage 1
trains a streaming ASR that later serves as both a probe and the
initialization. Stage 2 uses this model to probe a plausible commit
time for each character and applies a supervised warm-start; this
preserves recognition quality and yields an initial commit policy, but
the probed times are not treated as ground truth. Stage 3 refines the
commit policy with Group Relative Policy Optimization (GRPO). The
reward combines errors and latency, with errors taking priority, so
the model waits at ambiguities and commits once the evidence is
sufficient.
In summary, our contributions are as follows:

\begin{itemize}
\item We propose X2Streaming-ASR, a low-latency, high-accuracy Chinese--English streaming recognition system, which replaces global fixed delay with position-dependent commit decisions for adaptive character commitment.

\item We introduce a three-stage training framework: supervised learning establishes the ASR capability, commit-time probing warm-starts the commit policy, and character-level GRPO with commit-segment credit assignment directly optimizes recognition accuracy and emission latency.
\item Across ten Chinese and English test sets, X2Streaming-ASR attains the lowest mean leftover-wait latency, $32$--$109\,\mathrm{ms}$ on Chinese and $15$--$85\,\mathrm{ms}$ on English, while recognition accuracy remains comparable to existing systems.
\end{itemize}

\section{Method}
\begin{figure*}[t]
  \centering
  \includegraphics[width=0.9\linewidth]{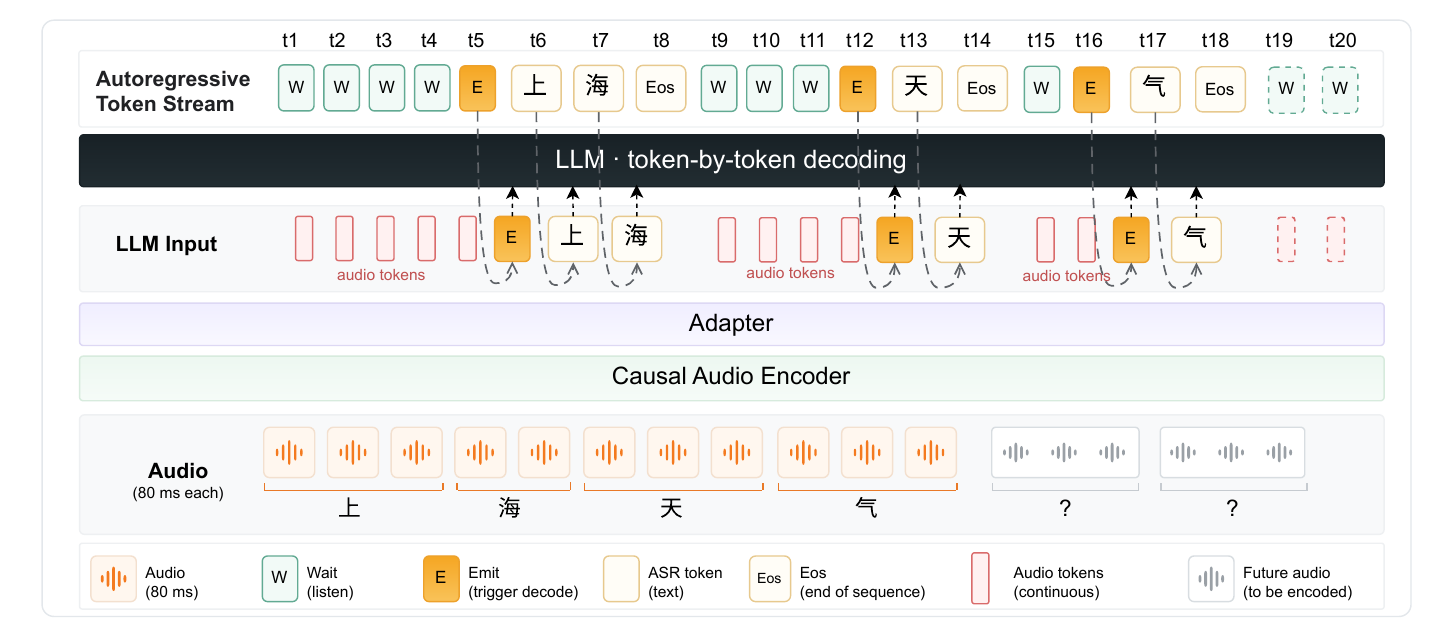}
  \caption{The X2Streaming-ASR model architecture alternates between listen and decode states. It's important to note that while the language model is decoding the previous audio segment, the causal audio encoder is simultaneously encoding the current audio; these two processes operate asynchronously.}
  \label{fig:X2Streaming-ASR}
\end{figure*}
\vspace{-0.5em}
\subsection{Architecture}
As shown in Fig~\ref{fig:X2Streaming-ASR}, X2Streaming-ASR is built on Voxtral Realtime, which consists of a causal audio encoder, an adapter, and a decoder-only language model.
Voxtral Realtime conditions the language model on a global delay $\tau$ and places text and audio tokens at the same sequence position.
X2Streaming-ASR drops $\tau$ and instead lets the model decide whether to wait or commit.

Given a 16kHz waveform, we extract a log-Mel spectrogram  and map it to continuous audio tokens $\mathbf{a}_{1:L}$ in the text embedding space.
The audio tokens have a frame rate of 12.5Hz, corresponding to 80ms of audio, which are interleaved with other tokens.
During streaming recognition, the language model alternates between listening and Decoding states. 
While Listening, the language model takes the historical interleaved sequence $z^{(t)}$ and the current audio token $a_t$ as input, where $z^{(t)}$ consists $a_{1:t}$, previously recognized text tokens and special tokens $e$.
The language model then makes a binary decision:
\begin{equation}
    c_t \sim p_\theta (c|z^{(t)}, a_t), \quad c \in\{w, e\}.
\end{equation}
Here $w$ means "wait", which is not written back; $e$ means "emit". If $c_t = w$, the language model stays in listening and reads next audio token $a_{t+1}$. If $c_t = e$, the language model will write $e$ back to the input as the starting point for recognition, stop reading new audio token, and perform autoregressive generates text token $y_k$ utils \(\langle\mathrm{Eos}\rangle\) token, where 
\begin{equation}
    y_k \sim p_\theta(y | z^{(t)}, a_t, e).
\end{equation}
Then X2Streaming-ASR return to listen state and \(\langle\mathrm{Eos}\rangle\) token is not written back into input.

\vspace{-0.5em}
\subsection{Training Strategy}
For Chinese the alignment and scoring unit is the character; for English it is the word.
The procedure below is described for characters and applies to English words in the same way.
To enable X2Streaming-ASR to decide both when to commit and what to commit, we use a three-stage training strategy.
Stage~1 trains a streaming recognizer that, given incremental audio, emits only characters whose acoustics have already ended; for example, two and a half characters of audio yield only the first two characters.
This checkpoint initializes Stage~2 and provides its labels.
We use Qwen3-ForcedAligner\cite{mu2026llm} to align the reference, then randomly cut the aligned character sequence into contiguous blocks of length $L \in \{1,\ldots,6\}$.
The commit time of a block is the end time of its last character plus $\Delta t \in \{0\,\mathrm{ms},\, 80\,\mathrm{ms},\, 160\,\mathrm{ms}\}$, and does not pass the midpoint of the next character if one exists:
\begin{equation}
    t_{\mathrm{emit}} = \min\bigl(t_{\mathrm{end}} + \Delta t,\; t^{\mathrm{next}}_{\mathrm{start}} + \tfrac{1}{2}\, d_{\mathrm{next}}\bigr).
\end{equation}
In Stage~1 we mask the loss on $w$ and $e$ and train only the ASR output.
Stage~2 labels each reference character's emit time from left to right with the Stage-1 model.
Probing starts at the end of the first character. At time $t$, the model greedily decodes the incremental audio.
Let $y_{1:k}$ be the longest reference prefix matched by the hypothesis; these $k$ characters share $t_{\mathrm{emit}}=t$.
For the next mismatched character $y_{k+1}$, probing moves to $t^{(k+1)}_{\mathrm{end}}$ if $t$ is earlier, and otherwise continues at $t+80\,\mathrm{ms}$, never backward.
An utterance is discarded if probing ends before every reference character receives a commit time, or if any character has $t_{\mathrm{emit}}-t_{\mathrm{end}}>640\,\mathrm{ms}$.
Such long waits teach the model to keep choosing $w$ and can prevent it from emitting $e$.
Stage~2 initializes from Stage~1 and trains recognition together with this initial commit policy by next-token supervision on the probed labels.
In Stages~1 and~2 we mix in offline full-text recognition at a ratio of $0.2$, so the same model can also recognize offline.
Stage~3 refines the commit policy with group-relative policy optimization (GRPO).
We sample a group of wait--emit trajectories \(\{\tau_k\}_{k=1}^{K}\) at the sentence level.
The action at each frame is \(c_t\in\{w,e\}\), drawn from \(\pi(\,\cdot\mid z^{(t)}, a_t)\).
If \(c_t=e\), the model decodes greedily until \(\langle\mathrm{Eos}\rangle\).
A commit segment is an $e$ together with the consecutive $w$ that precede it, so a trajectory splits into several commit segments.
The sentence is the sampling unit, but not the credit unit: broadcasting one sentence-level reward to every frame would reinforce useful and superfluous waits equally, and the policy would learn to wait throughout the utterance.
We therefore compare trajectories on the same reference character and assign the advantage to the corresponding commit segment.
Concretely, we align each hypothesis to the forced-aligned reference \(y_{1:J}\) and score character \(y_j\) on trajectory \(k\) by
\begin{equation}
s_{k,j}=
\begin{cases}
-\lambda\, d_{k,j}, & \text{correct alignment},\\
-1, & \text{substitution or deletion}.
\end{cases}
\end{equation}
The latency is normalized as
\begin{equation}
d_{k,j}=\frac{\min\bigl(\max(\ell_{k,j},0),H\bigr)}{H+1}\in[0,1),
\end{equation}
where \(\ell_{k,j}=t_{\mathrm{emit}}^{(k,j)}-t_{\mathrm{end}}^{(j)}\),
\(\lambda\) is a latency weight, and \(H\) is a truncation cap.
Subtracting the group mean on the same character gives
\begin{equation}
A_{k,j}=s_{k,j}-\bar{s}_{j},\qquad
\bar{s}_{j}=\frac{1}{K}\sum_{k=1}^{K}s_{k,j}.
\end{equation}
Each \(A_{k,j}\) is accumulated onto the commit segment that produced or missed that character.
Insertions have no reference character, so their penalty is assigned to the commit segment that produced them, followed by a zero-sum correction within the group.
Let the \(m\)-th commit segment of trajectory \(k\) be \(S_{k,m}\), the reference characters aligned to it be \(\mathcal{J}_{k,m}\), and the insertion penalty be \(A^{\mathrm{ins}}_{k,m}\).
The advantage of \(S_{k,m}\) is
\begin{equation}
\hat{A}_{k,m}
=
\sum_{j\in\mathcal{J}_{k,m}} A_{k,j}
+
A^{\mathrm{ins}}_{k,m}.
\end{equation}
Every decision in the segment shares \(\hat{A}_{k,m}\), so step \(t\) inherits \(\hat{A}_{k,m(t)}\) from the segment \(m(t)\) that contains it.
We optimize the current policy \(\pi_\theta\) against the sampling policy \(\pi_{\mathrm{old}}\) with the importance ratio
\begin{equation}
\rho_t
=
\frac{\pi_\theta(c_t\mid z^{(t)})}
{\pi_{\mathrm{old}}(c_t\mid z^{(t)})}.
\end{equation}
At decision step $t$ we maximize
\begin{equation}
\mathcal{L}
=
\mathbb{E}
\Biggl[
\min\Bigl(
\rho_t\hat{A}_{k,m(t)},\;
\mathrm{clip}(\rho_t,1-\varepsilon,1+\varepsilon)\,
\hat{A}_{k,m(t)}
\Bigr)
\Biggr],
\end{equation}
which strengthens or suppresses the corresponding $w/e$ actions according to the segment advantage, while keeping \(\rho_t\) near \(1\).
so that a single update cannot move too far.
\section{Experiments}
\begin{table*}[t]
  \centering
  \small
  \caption{Corpus CER/WER~(\%) and character/word-level \textbf{Mean} and \textbf{P95}
    latency~(ms). Chinese and English datasets report CER and WER,
    respectively. Offline results are shown where available.
    X2Streaming-ASR, X-ASR, and Zipformer use one checkpoint for streaming and
    offline recognition; Paraformer uses a separate offline model.
    Confucius4-R2T2 Speed and Quality are two streaming inference
    settings of the same model and share one offline result.
    \textbf{Uni-ASR streaming is the $1000\,\mathrm{ms}$ chunk,
    beam-$3$ result from the paper}$^\ddagger$.
    $^\ddagger$Not a matched-protocol comparison.
    Best / second-best streaming results are
    \textbf{bold} / \underline{underlined}.
    Best / second-best offline results are
    $^\star$ / $^\dagger$.}
  \label{tab:main_streaming}
  \setlength{\tabcolsep}{1.3pt}
  \renewcommand{\arraystretch}{1.08}

  \resizebox{\textwidth}{!}{%
  \begin{tabular}{l *{2}{cccc} ccc ccc c *{2}{cccc} cc}
    \toprule
    & \multicolumn{4}{c}{\textbf{X2Streaming-ASR}}
    & \multicolumn{4}{c}{\textbf{X-ASR}}
    & \multicolumn{7}{c}{\textbf{Confucius4-R2T2}}
    & \multicolumn{4}{c}{\textbf{Zipformer}}
    & \multicolumn{4}{c}{\textbf{Paraformer}}
    & \multicolumn{2}{c}{\textbf{Uni-ASR}$^\ddagger$}
    \\
    \cmidrule(lr){2-5}
    \cmidrule(lr){6-9}
    \cmidrule(lr){10-16}
    \cmidrule(lr){17-20}
    \cmidrule(lr){21-24}
    \cmidrule(lr){25-26}

    \textbf{Test Set}
    & \multicolumn{3}{c}{Streaming} & Off.
    & \multicolumn{3}{c}{Streaming} & Off.
    & \multicolumn{3}{c}{Str.(Speed)}
    & \multicolumn{3}{c}{Str.(Quality)} & Off.
    & \multicolumn{3}{c}{Streaming} & Off.
    & \multicolumn{3}{c}{Streaming} & Off.
    & Str. & Off. \\
    \midrule

    \textbf{Zh}
      & CER & Mean & P95 & CER
      & CER & Mean & P95 & CER
      & CER & Mean & P95
      & CER & Mean & P95
      & CER
      & CER & Mean & P95 & CER
      & CER & Mean & P95 & CER
      & CER & CER \\
    \midrule

    AISHELL-1
      & \textbf{1.53} & \textbf{32} & \textbf{320} & 0.89$^\star$
      & 4.81 & 651 & 880 & {3.01}
      & 3.55 & \underline{308} & \underline{560}
      & 2.29 & 735 & 1280
      & {1.69}
      & \underline{1.93} & 472 & 800 & 1.39$^\dagger$
      & 3.09 & 586 & 880 & 2.29
      & 2.15 & 1.44 \\

    AISHELL-2
      & {4.12} & \textbf{79} & \textbf{400} & {3.38}
      & {4.84} & {657} & {924} & {3.63}
      & {4.61} & \underline{319} & \underline{560}
      & \underline{3.58} & {627} & {1040}
      & 2.98$^\dagger$
      & 4.17 & 450 & 800 & 3.24
      & 3.92 & 568 & 840 & 3.02
      & \textbf{3.25} & 2.60$^\star$ \\

    AISHELL-3
      & \textbf{2.08} & \textbf{57} & \textbf{400} & 1.5$^\star$
      & 5.14 & 660 & 899 & {3.74}
      & 3.98 & \underline{321} & \underline{560}
      & \underline{2.85} & 672 & 1120
      & {2.37}
      & 2.95 & 465 & 800 & 2.17$^\dagger$
      & 3.72 & 576 & 849 & 2.76
      & {--} & {--} \\

    WS Meeting
      & 8.67 & \textbf{74} & \textbf{320} & {7.28}
      & 11.28 & 595 & 810 & {7.58}
      & 8.63 & \underline{320} & \underline{560}
      & \textbf{7.37} & 685 & 1200
      & {6.59}
      & \underline{7.88} & 436 & 800 & 6.24$^\star$
      & 10.06 & 582 & 880 & 6.98
      & 8.04 & 6.32$^\dagger$ \\

    WS Net
      & \underline{7.88} & \textbf{109} & \textbf{400} & {6.98}
      & 8.63 & 563 & 800 & 6.32$^\dagger$
      & 9.52 & \underline{310} & \underline{480}
      & 8.48 & 570 & 960
      & {8.1}
      & 8.53 & 409 & 720 & 6.91
      & 8.36 & 563 & 880 & 6.37
      & \textbf{6.44} & 5.78$^\star$ \\

    \midrule

    \textbf{En}
      & WER & Mean & P95 & WER
      & WER & Mean & P95 & WER
      & WER & Mean & P95 & WER
      & Mean & P95  & WER
      & WER & Mean & P95 & WER
      & WER & Mean & P95 & WER
      & WER & WER \\
    \midrule

    LS Clean
      & \underline{2.20} & \textbf{29} & \textbf{240} & {2.19}
      & 3.86 & 574 & 800 & {2.78}
      & 2.83 & \underline{194} & \underline{320}
      & \textbf{2.11} & 472 & 800
      & 1.8$^\star$
      & {--} & {--} & {--} & {--}
      & {--} & {--} & {--} & {--}
      & \underline{2.44} & 1.93$^\dagger$ \\

    LS Other
      & \underline{5.50} & \textbf{85} & \underline{480} & {4.91}
      & 9.64 & 585 & 800 & {6.22}
      & 6.38 & \underline{201} & \textbf{320}
      & \textbf{4.81} & 463 & 720
      & 3.70$ ^\star$
      & {--} & {--} & {--} & {--}
      & {--} & {--} & {--} & {--}
      & 5.71 & 4.11$^\dagger$ \\

    GS Clean
      & \textbf{10.01} & \textbf{43} & \textbf{320} & 9.28$^\star$
      & 10.47 & 592 & 847 & {10.07}
      & 10.69 & \underline{204} & \underline{400}
      & \underline{10.21} & 501 & 880
      & 9.92$^\dagger$
      & {--} & {--} & {--} & {--}
      & {--} & {--} & {--} & {--}
      & {--} & {--} \\

    VoxPopuli
      & \textbf{7.07} & \textbf{12} & \textbf{320} & 6.32$^\dagger$
      & 11.25 & 603 & 880 & {9.46}
      & 8.97 & \underline{204} & \underline{400}
      & \underline{7.26} & 523 & 880 & 6.28$^\star$
      & {--} & {--} & {--} & {--}
      & {--} & {--} & {--} & {--}
      & {--} & {--} \\

    TED-LIUM
      & \textbf{3.42} & \textbf{15} & \textbf{240} & 3.21$^\star$
      & 3.81 & 590 & 880 & {3.49}
      & 3.98 & \underline{196} & \underline{320}
      & \underline{3.45} & 497 & 880
      & 3.46$^\dagger$
      & {--} & {--} & {--} & {--}
      & {--} & {--} & {--} & {--}
      & {--} & {--} \\

    \bottomrule
  \end{tabular}}
\end{table*}
\begin{table}[t]
  \centering
  \small
  \caption{AISHELL-1 under forced post-boundary delay, natural wait/emit,
  and offline decoding with same checkpoint.
  A single emit may cover later characters, which then do not receive
  their own \(\Delta t\).}
  \label{tab:fixed_delay}
  \begin{tabular}{lccc}
    \toprule
    \textbf{Inference} & \textbf{CER(\%)} & \textbf{Mean lat.} & \textbf{P95} \\
    \midrule
    Forced \(t_{\mathrm{end}}\) & 7.46 & 0.07 & 0 \\
    Forced \(t_{\mathrm{end}}\)\(+80\,\mathrm{ms}\) & 3.48 & 80 & 80 \\
    Forced \(t_{\mathrm{end}}\)\(+240\,\mathrm{ms}\) & 2.01 & 214 & 240 \\
    Forced \(t_{\mathrm{end}}\)\(+640\,\mathrm{ms}\) & 1.58 & 401 & 640 \\
    Streaming & \textbf{1.53} & \textbf{32} & 320 \\
    \bottomrule
  \end{tabular}
\end{table}
\vspace{-0.5em}
\begin{table}[t]
  \centering
  \caption{AISHELL-1 listen-policy ablation.}
  \label{tab:ablate_s2}
  \setlength{\tabcolsep}{3.5pt}
  \begin{tabular}{llcccc}
    \toprule
    Model & Mode & CER & Mean lat.& P95 \\
    \midrule
    Proposed & Streaming & 1.53 & 32 & 320 \\
    Stage~2 & Streaming & 6.27 & 4 & 80 \\
    Stage~2 & Forced \(t_{\mathrm{end}}\) & 7.17 & 0.4 & 0 \\
    Stage~3 + content KL & Streaming & 2.37 & 46 & 160 \\
    Stage~3 + content KL & Forced \(t_{\mathrm{end}}\) & 7.16 & 0.4 & 0 \\
    Sentence-level reward & Streaming & 1.09 & 346 & 880 \\
    \bottomrule
  \end{tabular}
\end{table}
\vspace{-0.5em} 
\subsection{Setup}
For Chinese, we train Stage1 on AISHELL-1\cite{bu2017aishell1}, AISHELL-2\cite{du2018aishell2}, AISHELL-3\cite{shi2020aishell3}, AISHELL-4\cite{fu2021aishell4}, AliMeeting\cite{Yu2022M2MeT}, and WenetSpeech\cite{zhang2022wenetspeech}. For English, we train Stage1 on VoxPopuli\cite{wang2021voxpopuli}, TED-LIUM\cite{rousseau2012ted}, LibriSpeech\cite{panayotov2015librispeech}, and GigaSpeech\cite{chen2021gigaspeech}.
Stage~2 needs commit-time labels. Annotating the full training set is costly and unnecessary, because this stage only has to teach the model an initial commit decision. Therefore we probe 1/10 of the data.
Stage~3 uses the same data, so that supervised and RL listen policies
are compared on the same labeled pool.
Stage~3 rewards need only the reference characters and their
\(t_{\mathrm{end}}\), and do not require extra emit-time labels.
For Chinese, evaluation is on the AISHELL-1/2/3 test sets and the WenetSpeech meeting and net tests. For English, evaluation is on the LibriSpeech test-clean and test-other sets, GigaSpeech clean, VoxPopuli, and TED-LIUM.
For each test utterance, Qwen3-ForceAligner gives character end times
\(t_{\mathrm{end}}\); leftover wait is
\(\ell=t_{\mathrm{emit}}-t_{\mathrm{end}}\).
We report mean latency and P95 relative to the acoustic
boundary, and CER for recognition.
We use Zipformer, Paraformer, X-ASR\cite{xasr}, and Confucius4-R2T2\cite{confucius4r2t2} as open-source baselines under their public configurations. Uni-ASR~\cite{xia2026uni} is not open-sourced; we quote its paper CER (1000ms chunk, beam size 3) and cannot measure leftover-wait latency.
In Stage~3, \(K=8\) trajectories are sampled per utterance at
temperature \(1.4\), with \(\lambda=0.1\),
\(H=2000\,\mathrm{ms}\), and \(\varepsilon=0.2\).
Stages~1 and~2 train all parameters.
Stage~3 freezes the encoder and adapter and fine-tunes the language
model with LoRA.

\vspace{-0.5em} 
\subsection{Comparison with Baseline}
Table~\ref{tab:main_streaming} compares X2Streaming-ASR with open-source
streaming models and with Uni-ASR.
Overall, X2Streaming-ASR has the lowest mean leftover-wait latency,
$32$--$109\,\mathrm{ms}$ on Chinese and $12$--$85\,\mathrm{ms}$ on English.
Apart from Confucius4-R2T2 Speed, the open-source baselines mostly lie
around $400$--$740\,\mathrm{ms}$, with P95 latency typically from
$720\,\mathrm{ms}$ to more than $1\,\mathrm{s}$.
The P95 latency of X2Streaming-ASR is also modest, $240$--$480\,\mathrm{ms}$.
Confucius4-R2T2 Speed is the closest open-source system, at about
$310$--$320\,\mathrm{ms}$ mean latency on Chinese and
$190$--$200\,\mathrm{ms}$ on English, and it is still several times slower.

At this very low latency, X2Streaming-ASR remains in the first tier of
accuracy on most test sets.
On Chinese, it outperforms the open-source models and Uni-ASR on AISHELL-1 and AISHELL-3.
On WenetSpeech Net, its streaming CER is better than the other open-source models.
Uni-ASR reports a lower CER on that set, under a $1000\,\mathrm{ms}$ chunk
and beam size $3$, without leftover-wait latency.
The gap to open-source models is on AISHELL-2 and WenetSpeech Meeting,
where Confucius4-R2T2(Quality) is more accurate; on Meeting, Confucius4-R2T2(Speed) is more accurate as well.
On English, X2Streaming-ASR ranks first or second relative to the
open-source models on every set.
It surpasses Confucius4-R2T2(Speed) on every streaming WER and still has a
lower mean latency than Confucius4-R2T2(Speed).
X2Streaming-ASR achieves state-of-the-art performance on the GigaSpeech (clean subset), VoxPopuli, and TED-LIUM datasets; on the LibriSpeech test-clean and test-other sets, it ranks just behind Confucius4-R2T2 (Quality) with a negligible margin (except for test-other).

\vspace{-0.5em} 
\subsection{Global Post-Boundary Delay}
\begin{figure}[t]
  \centering
  \includegraphics[width=1.0\linewidth]{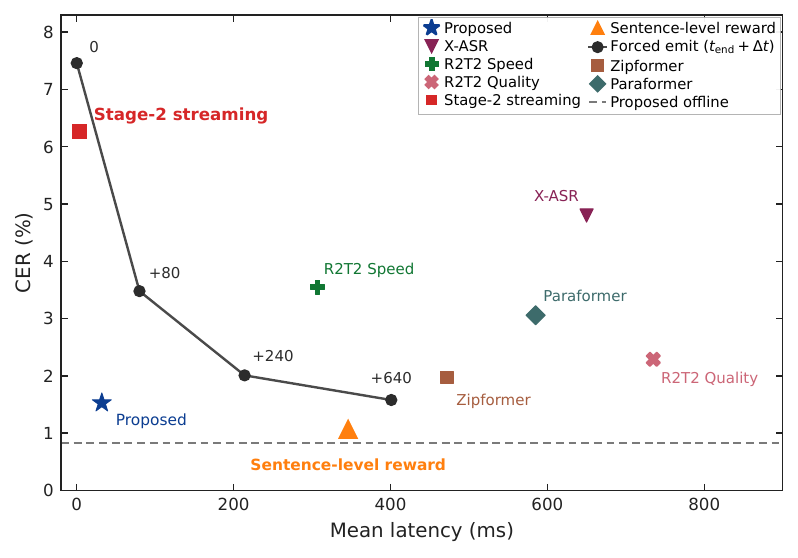}
  \caption{AISHELL-1 CER vs.\ mean character-level latency.
    The polyline is Forced emit at \(t_{\mathrm{end}}+\Delta t\)
    on the Stage-3 checkpoint.
    The dashed line is the same checkpoint with full-utterance decoding.}
  \label{fig:aishell1_cer_latency}
\end{figure}
To test whether X2Streaming-ASR has learned a single global wait,
we decode the AISHELL-1 test set with the same Stage-3 checkpoint
under forced post-boundary delays.
For reference characters not yet recognized, the model is forced to
start greedy decoding at
$t_{\mathrm{emit}}=t_{\mathrm{end}}+\Delta t$,
where $\Delta t \in \{0, 80, 240, 640\}\,\mathrm{ms}$.
A single emit may cover later characters, which then do not receive their own $\Delta t$; at large $\Delta t$, $t_{\mathrm{emit}}$ can also reach into the next character's acoustics.
Table~\ref{tab:fixed_delay} shows that committing at the character
endpoint is fastest ($0.07\,\mathrm{ms}$) but yields a CER of $7.46$.
An $80\,\mathrm{ms}$ wait lowers CER to $3.48$, still behind natural
streaming ($1.53$, $32\,\mathrm{ms}$).
A $240\,\mathrm{ms}$ wait matches that CER ($2.01$) only at
$214\,\mathrm{ms}$ mean latency, and $640\,\mathrm{ms}$ reaches $1.58$
at $401\,\mathrm{ms}$.
Streaming therefore matches the $640\,\mathrm{ms}$ wait at about
$13\times$ lower mean latency.
It is not selecting a compromise $\Delta t$;
it waits when uncertain and commits when certain.

\vspace{-0.5em} 
\subsection{Ablation Study}
To verify the impact of supervised learning on the submission decision, we evaluate Stage~2 on the AISHELL-1 test set.
As Table~\ref{tab:ablate_s2} and Figure~\ref{fig:aishell1_cer_latency} show, Stage~2 falls near the polyline and close to forced \(t_{\mathrm{end}}\), indicating that supervised learning tends to be global early.
To verify the benefit of the commit policy, we train Stage~3 with content-based KL divergence, anchoring recognition near Stage~2.
Rows 3 and 5 of Table~\ref{tab:ablate_s2} show that this KL term prevents recognition from drifting, and rows 1, 2, and 4 show that optimizing the commit policy is the main source of the CER and latency gains.
For comparison, we also train a GRPO variant with sentence-level rewards.
It lowers CER from $1.53$ to $1.09$, but mean latency rises from $32\,\mathrm{ms}$ to $346\,\mathrm{ms}$ and P95 from $320\,\mathrm{ms}$ to $880\,\mathrm{ms}$.
Sentence-level rewards therefore learn a global wait, sacrificing latency for better recognition.
\FloatBarrier

\section{Conclusion}
This paper proposes X2Streaming-ASR, a low-latency, high-accuracy
Chinese–English streaming recognition system, which separates when to commit from what to commit and trains recognition and the commit policy in stages, so the model waits when uncertain and commits when certain.
Experiments show that mean latency drops from hundreds of milliseconds to tens of milliseconds, while recognition accuracy stays in the first tier.
In the future, we will extend the method to more languages and realistic conditions to improve the model's robustness and recognition performance.


\bibliographystyle{IEEEbib}
\bibliography{refs}

\end{document}